\documentclass[debug,overfull]{epl}

\usepackage{graphics}
\usepackage{epsfig}
\usepackage{amsmath}
\usepackage{psfrag}

\title{Slow crack growth in polycarbonate films} \shorttitle{Slow crack growth in polycarbonate films}
\author{P.P. Cortet \and S. Santucci \and L. Vanel \and S. Ciliberto}
\shortauthor{P.P.Cortet \etal}

\institute{
  Laboratoire de physique, CNRS UMR 5672,
  Ecole Normale Sup\'erieure de Lyon,
  46 all\'ee d'Italie,
  69364 Lyon Cedex 07, France \\
}

\pacs{62.20.Mk}{Fatigue, brittleness, fracture and cracks}
\pacs{62.20.Fe}{Deformation and plasticity (including yield,
ductility, and superplasticity)} \pacs{05.70.Fh}{Phase
transitions: general studies}

\begin{document}

\maketitle

\begin{abstract}
We study experimentally the slow growth of a single crack in
polycarbonate films submitted to uniaxial and constant imposed
stress. The specificity of fracture in polycarbonate films is the
appearance of flame shaped macroscopic process zones at the tips
of the crack. Supported by an experimental study of the mechanical
properties of polycarbonate films, an analysis of the stress
dependence of the mean ratio between the process zone and crack
lengths, during the crack growth, show a quantitative agreement
with the Dugdale-Barenblatt model of the plastic process zone. We
find that the fracture growth curves obey strong scaling
properties that lead to a well defined growth master curve.

\end{abstract}

\section{Introduction}

Solids with a single crack usually break at a critical rupture
stress. However, experiments \cite{Zhurkov} show that a given
solid submitted to a subcritical stress breaks after a certain
amount of time. Therefore, understanding the mechanisms of
subcritical macroscopic fracture growth in solids has become an
important goal of fracture physics in order to improve the
resistance of structures to failure. Recent experimental works
\cite{Santucci1,Santucci2} have shown that sub-critical crack
growth in paper can be successfully described by a thermal
activation model for elastic brittle media. In this letter, we
present an experimental study of slow growth of a single crack in
a polycarbonate film which is a highly non-brittle material. We
observe that a large flame shaped area, the process zone, forms
ahead of each crack tip. We find that the dependence of the
process zone length with the applied stress, during the crack
growth, is in reasonable quantitative agreement with the
Dugdale-Barenblatt model. In that respect, polycarbonate appears
to be a good model material to understand the mechanisms of crack
growth in non-elastic media. We show that the crack growth curve
obeys remarkable scaling properties that are not theoretically
understood yet.

\section{The experimental setup and the experiment}

The experiment consists in loading $125 \mu$m thick isotropic
polycarbonate films (height $21$cm, length $24$cm) with uniaxial
and constant imposed stress $\sigma$. The polycarbonate films used
are made of Bayer Makrofol\textsuperscript{\textregistered} and
present the properties of bulk material. An initial crack of
length $\ell_i$ is initiated at the center of each polycarbonate
sample using calibrated blades of different lengths (from $0.5$cm
to $3$cm). Then, a constant force $F_c$ is applied to the film
perpendicularly to the crack direction, so that we get a mode 1
crack opening type. For more details about the setup see
\cite{Santucci1,Santucci2}. A high resolution and high speed
camera (Photron Ultima 1024) is used to follow the crack growth.

We follow the growth of the single linear fracture and its process
zones, under constant applied stress $\sigma$, till the total
rupture of the sample. The applied stress $\sigma$ is chosen such
that crack growth is slow, i.e., smaller than the critical one,
$\sigma_c$, above which a fast crack propagation occurs.

\section{Mechanical properties of polycarbonate films}

\begin{figure}
\onefigure[width=7cm]{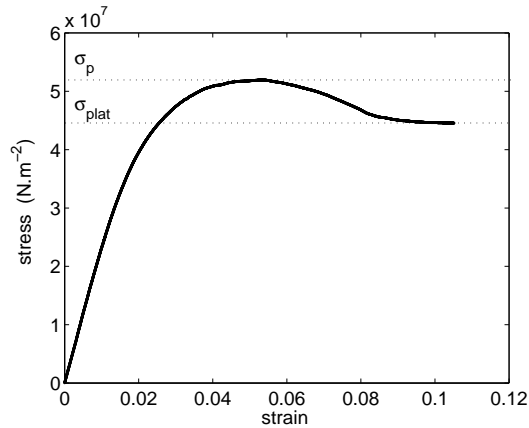} \caption{Stress as a function of
the strain for a $125\mu$m thick polycarbonate film loaded with a
$46.25 \mu \rm{m.s}^{-1}$ strain rate.} \label{rhéo}
\end{figure}

In order to characterize the material in which the crack will
grow, we performed some preliminary experiments on polycarbonate
films, without crack, submitted to uniaxial deformation at a
constant deformation rate ($46.25 \mu \rm{m.s}^{-1}$). A typical
experimental stress-strain plot is presented in figure \ref
{rhéo}. The polymer films show the classical behavior of a plastic
material with a quasi-elastic behavior for small strains followed
by a bell profile and a plateau. The different characteristic
values observed on this graph are in good agreement with the ones
measured by Lu and Ravi-Chandar in bulk polycarbonate \cite{Lu}.
We measured the experimental values of the maximum reachable
stress, $\sigma_p=5.2\,10^{7} \rm{N.m}^{-2}$, the plastic plateau
stress $\sigma_{plat}=4.45\,10^{7} \rm{N.m}^{-2}$ and the Young
modulus for small strains $Y=194\,10^7 \rm{N.m}^{-2}$.

\section{The flame shaped process zone ahead of the crack tip}

\begin{figure}
\psfrag{L}[c]{$\ell_{crack}$} \psfrag{H}[c]{$\ell_{pz}$}
\onefigure[width=12cm]{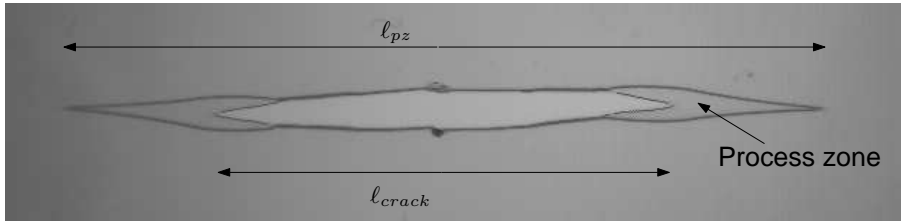} \caption{Image of a crack in a
polycarbonate film with its macroscopic process zone at each tip;
$\ell_{crack}$ is the crack length and $\ell_{pz}$ is the process
zone length from tip to tip.} \label{imfrac}
\end{figure}

\begin{figure}
    \centerline{
    \includegraphics[width=7cm]{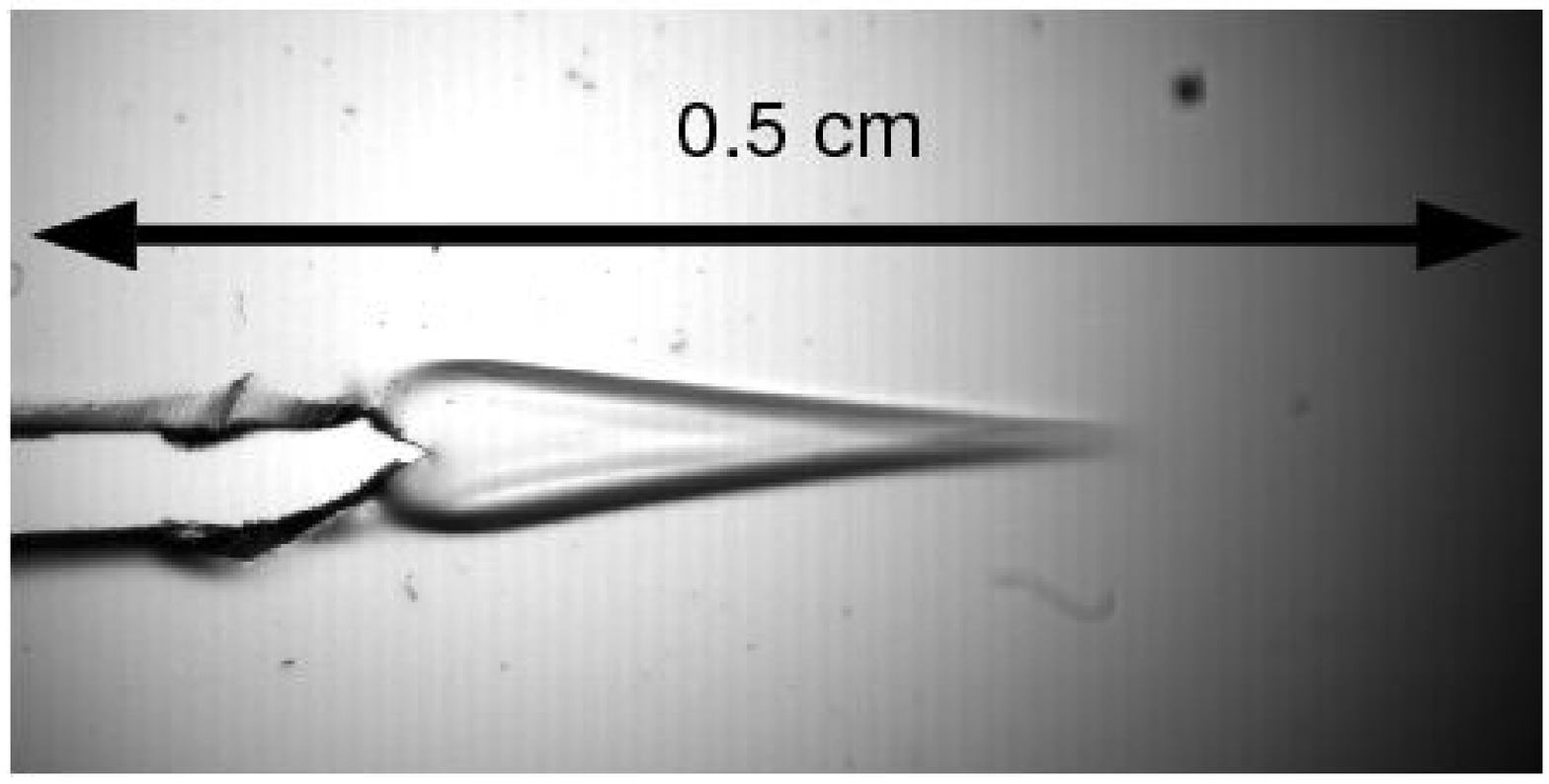}
    \includegraphics[width=4.67cm]{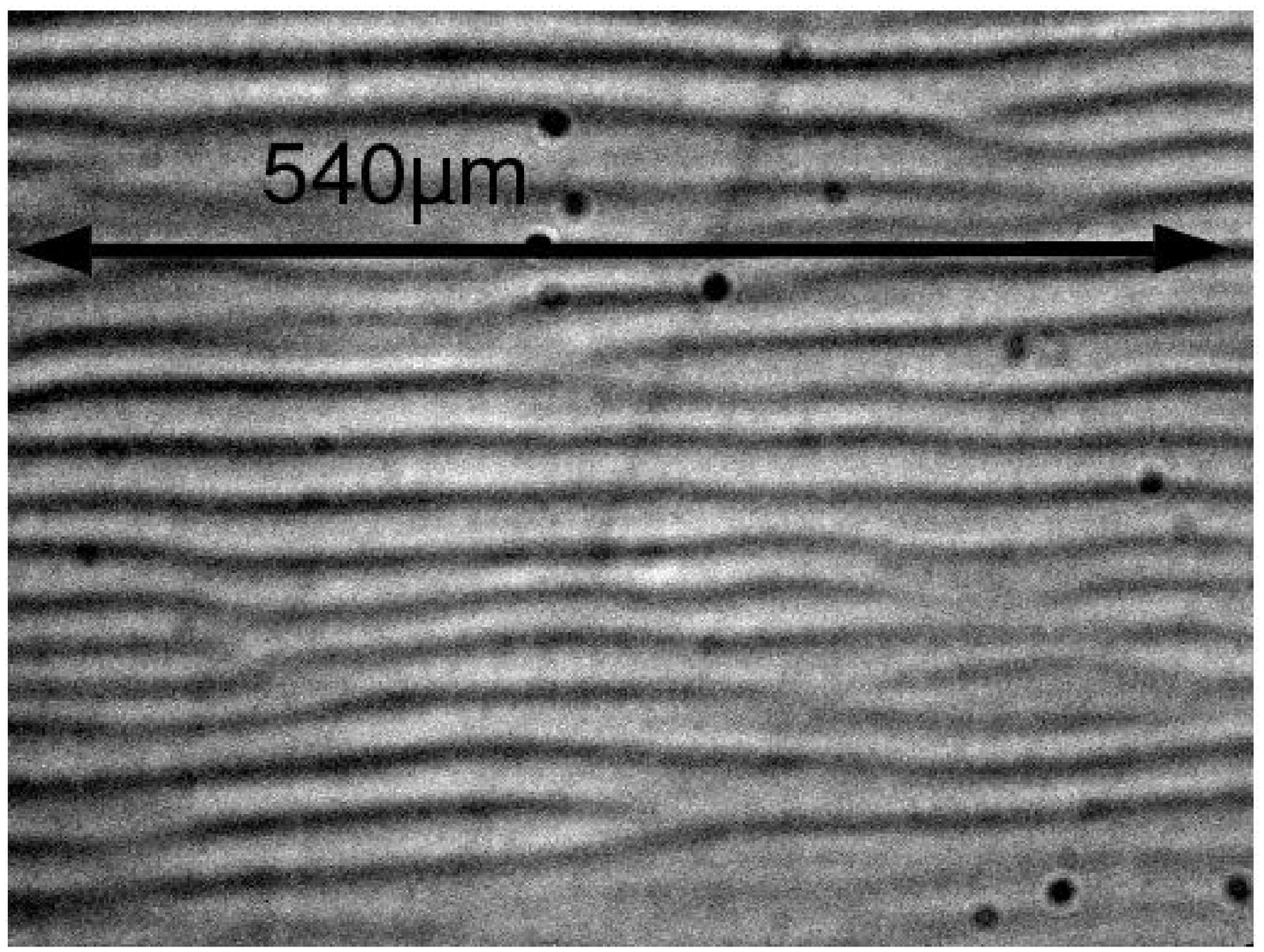}}
    \caption{On the left: process zone at the tip of a growing crack
    in a polycarbonate film; on the right: microscopic image showing
    striations quasi-parallel to the fracture direction in the process zone.}
    \label{flamme}
\end{figure}

In each experiment, during the loading phase of the film, a
macroscopic flame shaped process zone appears at each tip of the
crack and grows with the applied stress (cf. figures \ref{imfrac}
and \ref{flamme}). This zone was previously noticed by Donald and
Kramer \cite{Donald}. In the late loading stage, the fracture may
also grow a little. Consequently, the real experimental initial
condition, obtained when the constant stress $\sigma$ is reached,
is not exactly $\ell_i$. During the imposed stress stage, the
process zone and the fracture are both growing till the final
breakdown of the sample in a way that the fracture never catches
up the process zone tip.

Inside the process zone, the film is subjected to a thinning which
brings its thickness from $125 \mu$m to about $70 \mu$m (measured
on post-mortem samples). It is worth noticing that on microscopic
images (cf. figure \ref{flamme}) one can see in the process zone
the presence of striations quasi-parallel to the fracture front
with a wave length of about $22 \mu$m. These striations seem to be
thickness oscillations of the film. It is still an open question
whether this process zone, once it has been formed, continues to
behave as a visco-plastic zone or as an elastic zone with an
effective macroscopic Young modulus different from the one of the
rest of the film. Understanding the mechanical nature of the
process zone is the key to reach a model of fracture growth in
polycarbonate.

\section{Dependence of the process zone length on the fracture length}

\begin{figure}
    \psfrag{X}[c]{$\ell_{crack}$(cm)}
    \psfrag{Y}[c]{$\ell_{pz}$(cm)}
    \psfrag{P}[c]{$\ell_{pz}/\ell_{crack}$}
    \psfrag{Z}[l]{$\ell_i=3$cm $F_c=750$N}
    \psfrag{V}[l]{$\ell_i=2$cm $F_c=850$N}
    \psfrag{U}[l]{$\ell_i=1.5$cm $F_c=900$N}
    \psfrag{Q}[l]{$\ell_i=0.5$cm $F_c=1000$N}
    \leftline{\rm{a)} \hspace{6.5cm} \rm{b)}}
    \centerline{
    \includegraphics[width=6.80cm]{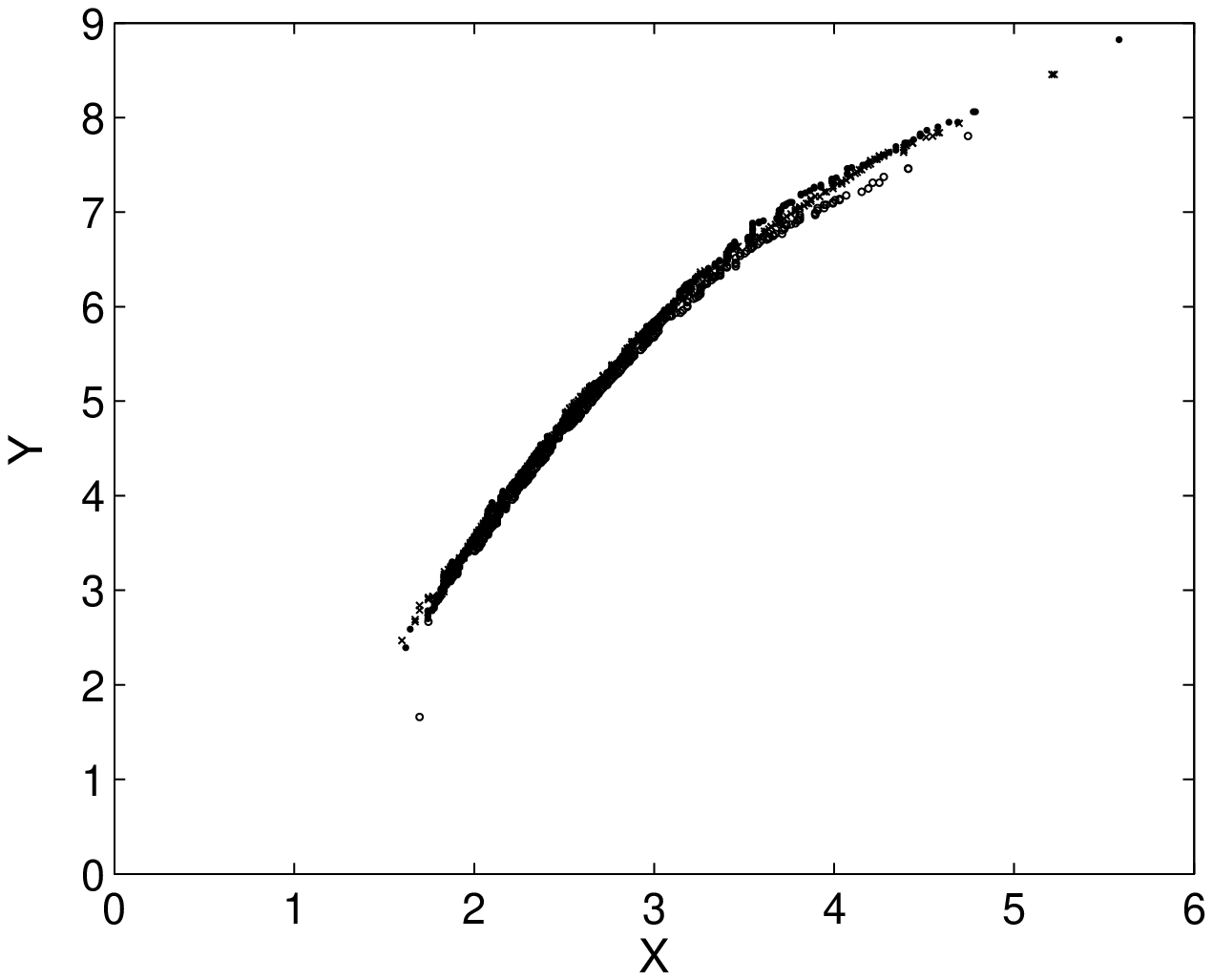}
    \includegraphics[width=7.1cm]{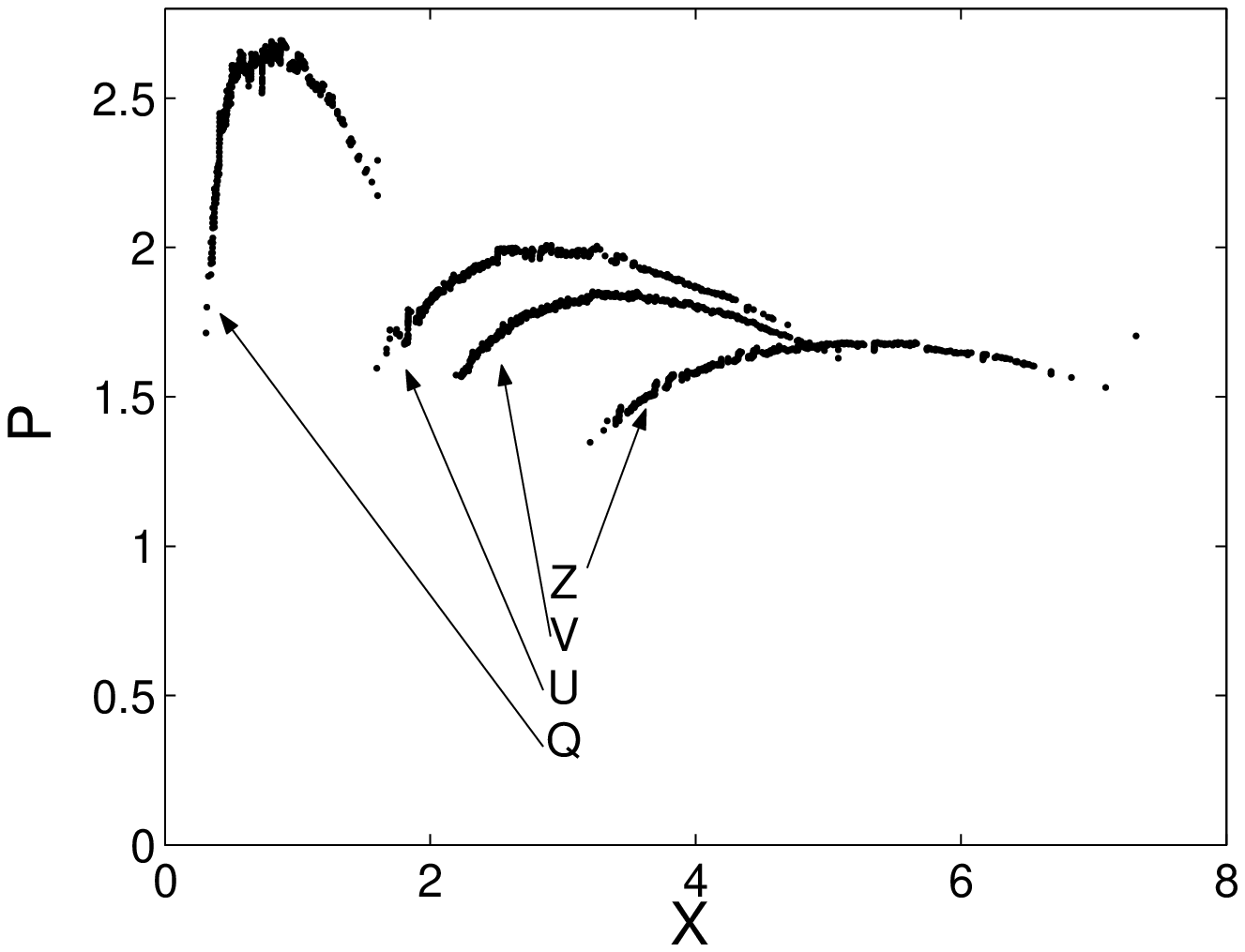}}
    \caption{On the left: process zone length as a function of the
    crack length for three experiments performed in the same experimental
    conditions ($\ell_i=1.5$cm and $F_c=900$N); on the right: $\ell_{pz}/\ell_{crack}$
    ratio for four experiments performed with different experimental conditions
    as a function of the crack length.}
    \label{Lp_Lf}
\end{figure}

The experimental relation between the process zone length (defined
on figure \ref{imfrac}), $\ell_{pz}$, and the crack length,
$\ell_{crack}$, during the crack growth process, is plotted on
figure \ref{Lp_Lf}a. It seems that this relation is not
statistical because all curves for identical experimental
conditions are almost identical. The experimental stress ranges
for different $\ell_i$ do not overlap, so that it is not possible
to plot the curves for the same $\ell_i$ and very different
$\sigma$. However, it is expected that the $\ell_{pz}$ against
$\ell_{crack}$ curve depends on the applied stress only.

It is important to notice that during an experiment the fracture
length grows on time scales much larger than the time scale of
relaxation of the process zone shape. This has been checked by
suddenly forcing the fracture length to grow using a blade during
an imposed stress experiment. This action induces a
quasi-instantaneous growth of the process zone to its new
equilibrium state. So, it is appropriate to think that during the
fracture experiments, the process zone is in a quasi-equilibrium
state for each given crack length (below the critical one) and
applied load. Therefore, we can analyze these data using an
equilibrium model for the process zone.

The relation between $\ell_{pz}$ and $\ell_{crack}$ has first been
theoretically described by the Dugdale-Barenblatt model
\cite{Dugdale,Barenblatt}. Dugdale considers the process zone as
an isotropic plastic material in which the stress is uniformly
equal to the plastic yield stress $\sigma_y$. Assuming the
non-divergence of the stress at the tip of the process zone, he
concludes to a zero stress intensity factor (SIF) \cite{Sif} at
the tip of the process zone:
$K_{tot}=K_{el}(\sigma;\ell_{pz})+\tilde{K}(\sigma_{y};\ell,\ell_{pz})=0$,
where $K_{el}$ is the traditional SIF at the tip of a fracture of
length $\ell_{pz}$ in an elastic film submitted to $\sigma$ at its
border and $\tilde{K}$ the SIF for a film fractured on a length
$\ell_{pz}$ and submitted only to $\sigma_y$ on the fracture lips
between $\ell_{crack}$ and $\ell_{pz}$. Using analytical
expressions for these SIFs in the case of an infinite elastic
sample, Dugdale finds a proportionnality dependence of $\ell_{pz}$
on $\ell_{crack}$: $\ell_{pz}/\ell_{crack}=
1/\cos\left(\frac{\pi}{2}\sigma/\sigma_y\right)$. This model was
successfully compared with experimental data in metals
\cite{Dugdale}. However, for many polymers, it does not predict
the correct order of magnitude for the process zone size if one
uses the yield stress constant of the material as the Dugdale
stress constant \cite{Stojimirovic,Haddaoui,Chud_length}. On
figure \ref{Lp_Lf}b, it appears clearly that the
$\ell_{pz}/\ell_{crack}$ ratio is not constant during a constant
stress experiment, so that the $\ell_{pz}$ against $\ell_{crack}$
relation is not a proportionality relation.

\begin{figure}
 \psfrag{X}[c]{$\sigma$ ($\,10^{7}\rm{N.m}^{-2}$)}
 \psfrag{Y}[c]{$\ell_{pz}/\ell_{crack}$}
 \psfrag{Z}[l][][0.8]{$1/\cos\left(\frac{\pi}{2}\frac{\sigma}{5.2\,10^7}\right)$}
\onefigure[width=7cm]{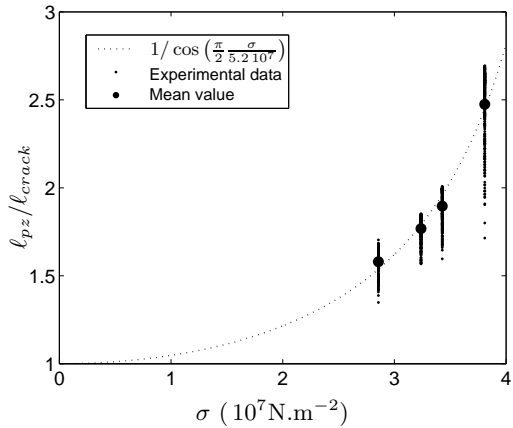} \caption{$\ell_{pz}/\ell_{crack}$
ratio as a function of the applied stress $\sigma$ for four
experiments performed with different experimental conditions and
the fit of its mean values (the larger points correspond to the
mean value of the data for each $\sigma$) by the
Dugdale-Barenblatt law.} \label{db}
\end{figure}

To make a more quantitative comparison with the model, we
represent, on figure \ref{db}, the $\ell_{pz}/\ell_{crack}$ ratio
as a function of the applied stress $\sigma$ for four experiments
performed with different applied stresses. The dispersion of the
data for each experiment corresponds to the important discrepancy
from a proportionality law between the two lengths. For each of
the four experiments, the mean value of the
$\ell_{pz}/\ell_{crack}$ ratio is superimposed on the data.
Fitting these mean values against $\sigma$ with the
Dugdale-Barenblatt law, $\ell_{pz}/\ell_{crack}=
1/\cos\left(\frac{\pi}{2}\sigma/\sigma_y\right)$, leads to an
estimate of the Dugdale stress constant: $\sigma_y=5.2\,10^{7}
\rm{N.m}^{-2}$. It is striking how the value found here for
$\sigma_y$ is close to the plastic peak stress $\sigma_{p}$ of the
rheology curve of figure \ref {rhéo}. So, even if the
Dugdale-Barenblatt model does not predict the non-linear
dependence of $\ell_{pz}$ on $\ell_{crack}$, it predicts the
correct order of magnitude for  the ratio
$\ell_{pz}/\ell_{crack}$, as a function of $\sigma$, using a
plastic stress constant $\sigma_y$ that corresponds very well to
the maximum reachable stress $\sigma_p$ obtained from mechanical
tests (cf. figure \ref {rhéo}). We insist that it is not usually
the case for polymers. We believe that the reason it works
reasonably well is that, for polycarbonate, plasticity remains
confined close to the crack tip while for many other polymers
plasticity is much more diffuse.

Regarding the non-proportionality of the curves in figure
\ref{Lp_Lf}a, finite size corrections to the Dugdale-Barenblatt
model lead to an opposite curvature \cite{Chud_length} to the one
observed experimentally. An explanation for the experimental
curvature, at the beginning or at the end of the experiment, could
be dynamical effects that bring the process zone out of
equilibrium. However, for the main part of the data, crack growth
is slow and process zone equilibrium models need to be improved to
explain the observed non-linearity.

\section{Breaking time and crack growth curves}

The measured breaking times for a given experimental condition
($\ell_i$, $F_c$) are statistical. Actually, we commonly observe a
factor five between the smallest and the largest breaking times.

\begin{figure}
    \psfrag{Z}[c]{$\ell_{crack}-\ell_{x}$(cm)}
    \psfrag{X}{$t$(s)}
    \psfrag{Y}{$t/\tau$}
    \psfrag{L}[c]{$\ell$}
    \psfrag{U}[c]{length(cm)}
    \leftline{\rm{a)} \hspace{6.5cm} \rm{b)}}
    \centerline{
    \includegraphics[width=6.84cm]{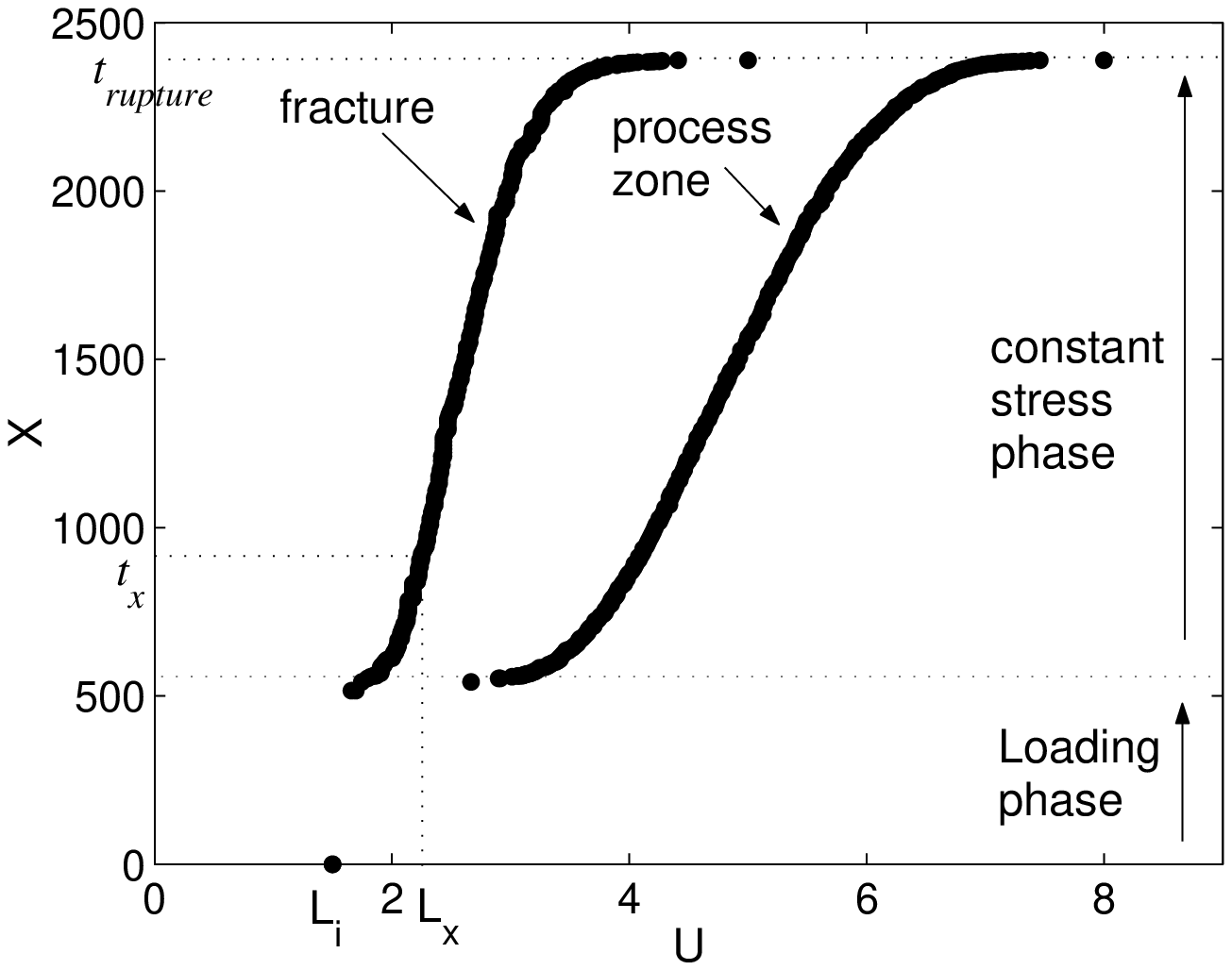}
    \includegraphics[width=7cm]{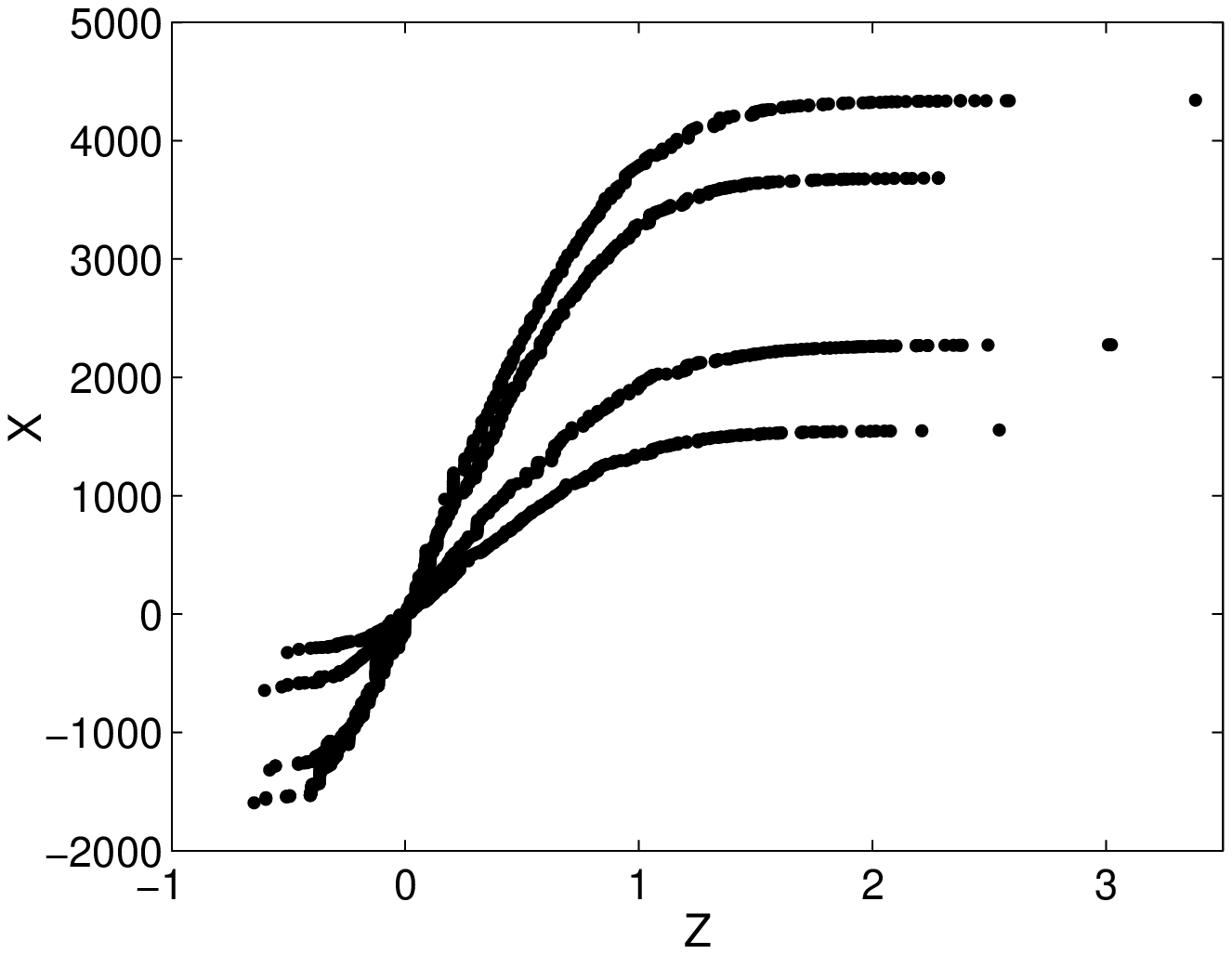}}
    \caption{On the left: time as a function of both the crack and process
    zone lengths for an imposed stress experiment ($\ell_i=1.5$cm, $F_c=900$N);
    on the right: time as a function of $\ell_{crack}-\ell_{x}$ for four experiments
    performed in the same experimental conditions ($\ell_i=1.5$cm,
    $Fc=900$N) with a change in time and length origins.}
    \label{t_long}
\end{figure}

Typical growth curves of the fracture and process zone are shown
on figure \ref{t_long}a. Both curves show a quite similar smooth
shape. We observe, once the loading phase is finished, at the
beginning of the constant stress phase, large velocities of the
fracture and the process zone tips which decrease till reaching
quasi-constant values before increasing back dramatically till the
final rupture. So, the first part of the growth curves in the
constant stress phase corresponds to a deceleration of the
fracture tips. We think this phenomenon is caused by delays in
deformation accumulated during the loading phase. In other words,
this deceleration corresponds to the time needed by the viscous
effects to relax all those delays to quasi-equilibrium. The final
acceleration corresponds to a transition to fast crack dynamics.

\begin{figure}
    \psfrag{X}[c]{$\ell_{crack}-\ell_{x}$(cm)}
    \psfrag{Z}[l][][0.7]{$t/\tau=1-\text{exp}[-1.3\,(\ell_{crack}-\ell_{x})]$}
    \psfrag{Y}[l][][0.7]{$t/\tau=\text{erf}[1.07\,(\ell_{crack}-\ell_{x})]$}
    \psfrag{U}{$t/\tau$}
    \leftline{\rm{a)} \hspace{6.5cm} \rm{b)}}
    \centerline{
    \includegraphics[width=7cm]{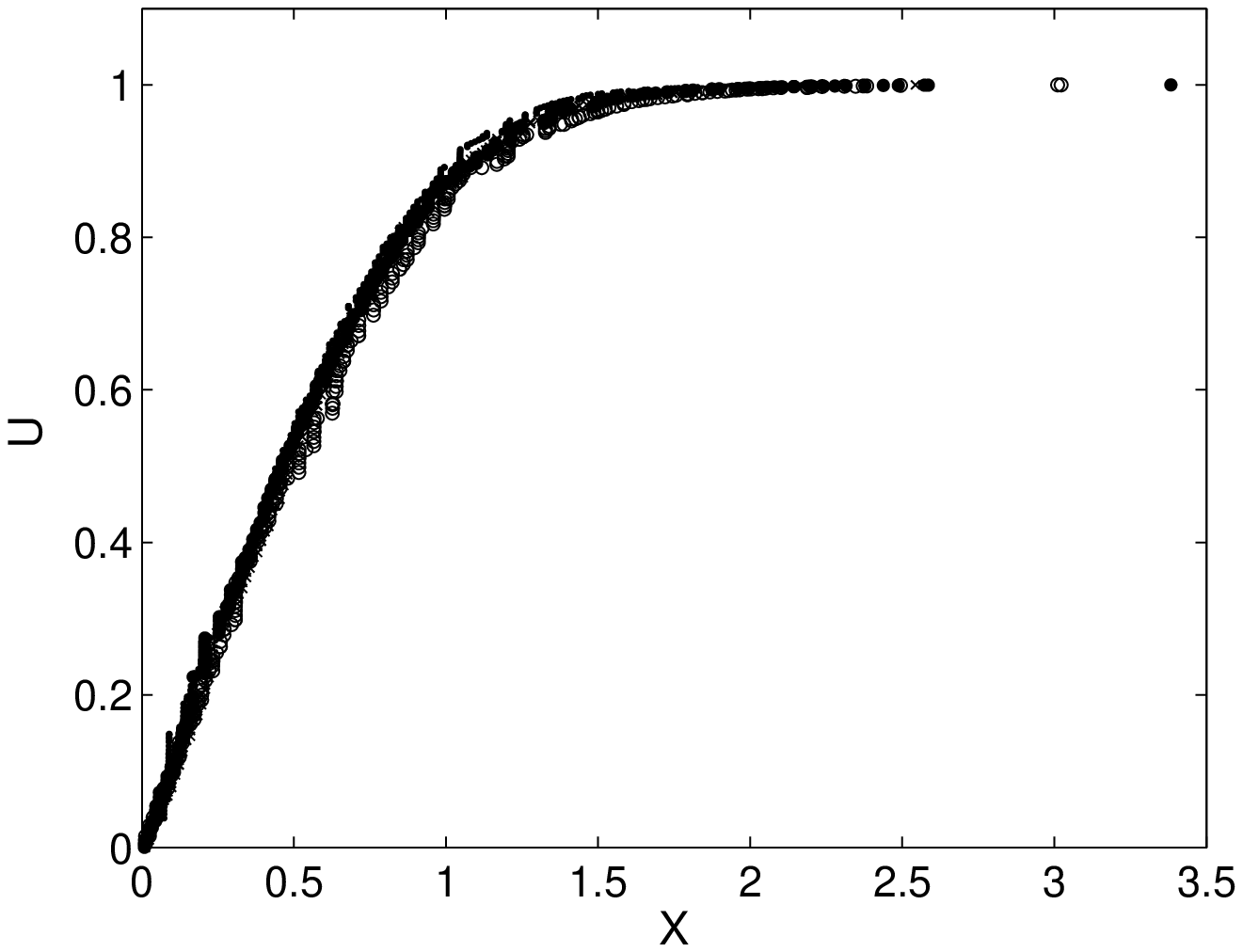}
    \includegraphics[width=6.96cm]{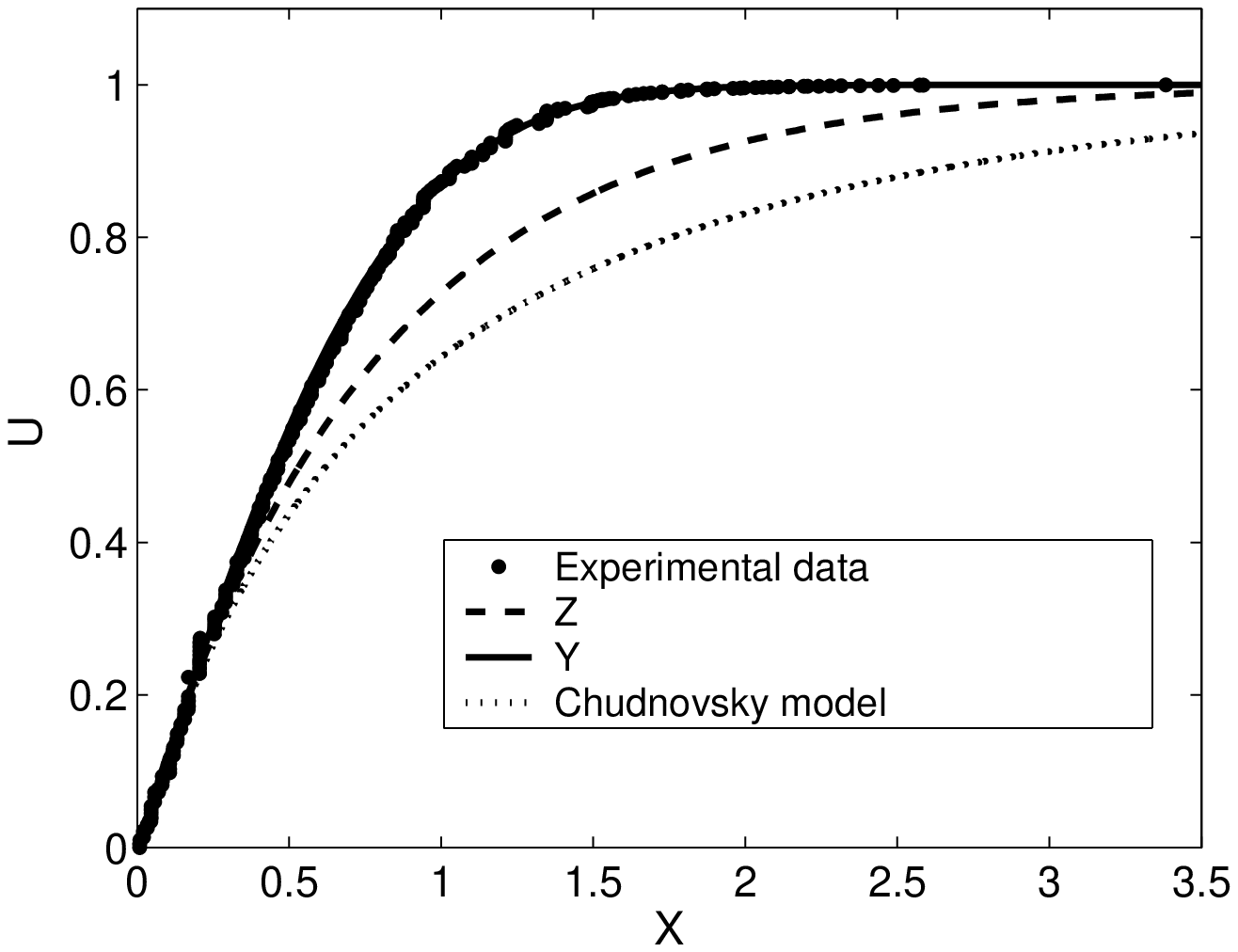}}
    \caption{On the left: time rescaled by the time left to rupture
    as a function of $\ell_{crack}-\ell_{x}$ for four experiments performed in
    the same experimental conditions ($\ell_i=1.5$cm, $F_c=900$N, same
    experiments as for figure \ref{t_long}b) with a change in time and
    length origins; on the right: time rescaled by the time left to
    rupture as a function of $\ell_{crack}-\ell_{x}$ for an imposed stress
    experiment ($\ell_i=1.5$cm, $Fc=900$N) and fits by the
    Chudnovsky model, the thermal activation model and an error function
    adjusting the initial velocity.}
    \label{t_long_fit}
\end{figure}

We can point out some important scaling properties of these growth
curves. Actually, by redefining the origin of time and length
using the coordinates of the inflexion point ($\ell_x$ and
$t_{x}$) of the fracture curve (cf. figure \ref{t_long}b) and
rescaling the time by the time left to break $\tau$
($\tau=t_{rupture}-t_x$), all the growth curves corresponding to
identical experimental conditions fall on a master curve (cf.
figure \ref{t_long_fit}a). By introducing a rescaling factor on
the lengths, we can as well make all growth curves, for different
experimental conditions, fall on the same master curve. This
rescaling property implies that the dynamics of the final part
(the part after the inflexion point) of the crack growth follows:

\begin{equation}
\label{rescaling} t=\tau(\sigma, \ell_i, \text{realization})\,
g\left(\frac{\ell- \ell_{x}}{\lambda}\right)
\end{equation}
where $g$ is the functional form of the master curve. $\lambda$
and $\ell_x$ are a priori functions of $\sigma$ and $\ell_i$. In
this formula, all the statistics is included in $\tau$. No
conclusive study of $\lambda(\ell_i,\sigma)$ has been possible due
to the limited ranges of $\ell_i$ and $F_c$ experimentally
accessible.

In searching for an analytical expression for $g$, we first tried
to use previously established models. Santucci et al.
\cite{Santucci1,Santucci3} developed a thermal activation model
that can explain sub-critical single crack growth in elastic
brittle films. This model that has been successfully faced with
experimental data for slow crack growth in paper
\cite{Santucci1,Santucci2}, predicts an exponential growth law: $t
= \tau \left[1 - \exp(-\frac{\ell-\ell_i}{\xi})\right]$. On figure
\ref{t_long_fit}b, we can see that this thermal activation law
(dashed line) does not work with experimental data for
polycarbonate. Obviously, it is not a surprise since polycarbonate
is far from being simply an elastic medium at least because of the
macroscopic plastic process zone. There are very few models giving
an equation of motion for crack growth in viscoplastic media. A
tentative model is the one of Chudnovsky \cite{Chud_growth} that
has been developed first for stick-slip crack growth description
in polyethylene films. This model introduces an aging process to
describe the creep of the process zone. By using the Dugdale
relation for process zone size, Chudnovsky reached a continuous
expression for the crack velocity as a function of the elastic
stress intensity factor. The curve obtained by a numerical
integration of this equation is shown on figure \ref{t_long_fit}b
(dotted line). It also does not fit the experimental data well.
Without any theoretical arguments, it appears that the error
function is a quite good guess candidate for an analytical
description of $g$ as we can see on the data fit of figure
\ref{t_long_fit}b (solid line).

\section{Conclusion}
The experimental study of the process zone of polycarbonate film
cracks allowed us to analyze the relevance of the
Dugdale-Barenblatt model. We have shown that the mean ratio
between the process zone and crack lengths, as a function of the
stress, agrees with this model. We have also found a
characteristic growth curve for crack in polycarbonate films which
cannot be explained by recent models based either on thermally
activated rupture or on surface energy aging due to viscoplastic
flow. It is important to introduce a more precise description of
the rupture mechanisms in the process zone in order to model crack
growth in polycarbonate. One direction is to consider the general
framework of viscoplastic rupture introduced by Kaminskii
\cite{Kaminskii1, Kaminskii2}. Work is in progress to test the
applicability of this model to crack growth in polycarbonate.


\begin{thebibliography}{0}

\bibitem{Zhurkov}
  \Name{Zhurkov S. N.}
  \REVIEW{Int. J. Frac. Mech.}{1}{1965}{311}.

\bibitem{Santucci1}
  \Name{Santucci S., Vanel L., Ciliberto S.}
  \REVIEW{Phys. Rev. Lett.}{93}{2004}{095505}.

\bibitem{Santucci2}
  \Name{Santucci S., Cortet P. P., Vanel L., Ciliberto S.}
  \text{to appear}.

\bibitem{Lu}
  \Name{Lu J., Ravi-Chandar K.}
  \REVIEW{Int. J. Solids. Structures}{36}{1999}{391}.

\bibitem{Donald}
  \Name{Donald A. M., Kramer E. J.}
  \REVIEW{J. Mater. Sci.}{16}{1981}{2967}.

\bibitem{Dugdale}
  \Name{Dugdale D. S.}
  \REVIEW{J. Mech. Phys. Solids}{8}{1960}{100}.

\bibitem{Barenblatt}
  \Name{Barenblatt G. I.}
  \REVIEW{Adv. Appl. Mech.}{7}{1962}{55}.

\bibitem{Sif}
  At a distance $r$ from a crack tip, the
  local stress behaves as $K /\sqrt{r}$,
  where $K$ is the stress intensity factor.

\bibitem{Stojimirovic}
  \Name{Stojimirovic A., Kadota K., Chudnovsky A.}
  \REVIEW{J. Appl. Polym. Sci.}{46}{1992}{1051}.

\bibitem{Haddaoui}
  \Name{Haddaoui N., Chudnovsky A., Moet A.}
  \REVIEW{Polymer}{27}{1986}{1377}.

\bibitem{Chud_length}
  \Name{Stojimirovic A., Chudnovsky A.}
  \REVIEW{Int. J. Frac.}{57}{1992}{281}.

\bibitem{Santucci3}
   \Name{Santucci S. $\etal$}
   \REVIEW{Europhys. Lett.}{62}{2003}{320}.

\bibitem{Chud_growth}
  \Name{Chudnovsky A., Shulkin Y.}
  \REVIEW{Int. J. Frac.}{97}{1999}{83}.

\bibitem{Kaminskii1}
  \Name{Kaminskii A. A.}
  \REVIEW{Sov. Appl. Mech.}{15}{1979}{1078}.

\bibitem{Kaminskii2}
  \Name{Kaminskii A. A.}
  \REVIEW{Int. Appl. Mech.}{40}{2004}{829}.

\end{thebibliography}
\end{document}